\documentclass{PoS}
\usepackage{multirow}
\title{Top-quark mass determination at the LHC:\\ a theory overview}
\newcommand{\DeltaMO}{\Delta_O^m}

\ShortTitle{Top-quark mass determination at the LHC: a theory overview}

\author{\speaker{Gennaro Corcella}\\
  INFN, Laboratori Nazionali di Frascati\\
  Via E.~Fermi 40, 00044 Frascati (RM), Italy\\
        E-mail: \email{gennaro.corcella@lnf.infn.it}}

\abstract{I briefly overview the methods employed at the LHC to extract
  the top-quark mass, taking particular care about the theory uncertainty
  and the dependence on the Monte Carlo hadronization parameters.}

\FullConference{The European Physical Society Conference on High Energy Physics
  \\		5-12 July, 2017\\
		Venice}

\begin{document}
\section{Introduction}
The mass of the top quark is a fundamental parameter of the
Standard Model, since it enters in the electroweak precision tests
\cite{gfitter} and constrained the mass of the Higgs boson 
even before its actual discovery at the LHC.
Moreover, the fact that the 
electroweak 
vacuum lies on the boundary
between stability and metastability regimes \cite{degrassi}
depends on the actual values of top and Higgs masses.
This statement does however depend on 
the identification of  
the top-quark mass world average, i.e.
$m_t=[173.34\pm 0.27{\rm (stat)} \pm 0.71{\rm (syst)}]$~GeV \cite{wave},
with the pole mass
and no extra uncertainty is included in the exploration of
Ref.~\cite{degrassi}.
In fact, any change of the central value or of the error on $m_t$
may affect the results in \cite{degrassi}, to the point
of even moving the vacuum position inside the stability or
instability regions.
It is therefore of paramount importance determining $m_t$ at the LHC
with the highest possible precision and, above all, estimating reliably
all sources of uncertainty.

The top-quark mass is determined
by comparing experimental data
with theory predictions: the extracted mass is the quantity 
$m_t$ in the calculation or in the Monte
Carlo event generator employed to simulate top production and decay.
In the following, I shall review the main methods used to reconstruct
the top-quark mass at the LHC and discuss the 
theoretical and Monte Carlo uncertainties, paying special attention
to the dependence on the event-generator $b$-fragmentation parameters.
I shall finally make some concluding remarks.

\section{Top-quark mass extraction at LHC}

Top-quark mass determinations at hadron colliders
are classified
as standard or
alternative measurements.
Standard top-mass analyses adopt the template, matrix-element
and ideogram methods (see, e.g., the analyses in \cite{atlas1,cms1}) and
compare final-state distributions, associated with
top-decay ($t\to bW$) products, such as the $b$-jet+lepton invariant
mass in the dilepton channel, with the predictions yielded by
the Monte Carlo codes.
Event generators like
the general-purpose HERWIG \cite{herwig} or PYTHIA \cite{pythia}
simulate the hard-scattering process 
at leading order (LO),
multi-parton emissions in the soft or collinear
approximation and the interference between
top-production and decay stages is neglected (narrow-width
approximation).
More recent NLO+shower programs, such as 
MadGraph5$\_$aMC@NLO \cite{mcnlo} and POWHEG \cite{powheg}, 
implement NLO
hard-scattering amplitudes, but still depend on 
HERWIG and PYTHIA for parton cascades and non-perturbative
phenomena, such as hadronization or underlying event.
As a whole, standard top-quark mass determinations, as they are
based on the reconstruction of the invariant mass
of the top-decay products and rely on programs which
factorize top production ad decay, should lead to 
results close to the top-quark pole mass.
However, as will be pointed out hereafter, 
a careful determination of the theoretical uncertainty,
of both perturbative and non-perturbative origins, such as
missing higher orders, width corrections and
colour-reconnection effects, is compelling.

Other strategies to measure $m_t$, making use of 
total or differential
cross sections, endpoints, energy peaks or kinematic
properties of $t\bar t$ final states,
are traditionally called `alternative' measurements.
The total $t\bar t$
cross section was calculated in the NNLO+NNLL approximation \cite{alex}
and allows a
direct determination of the pole mass  \cite{sigmaatl,sigmacms},
the mass definition used in the computation \cite{alex}.
The errors in \cite{sigmaatl} and \cite{sigmacms}
are larger than those in the standard methods;
however, they are expected to decrease thanks to the higher
statistics foreseen at the LHC Run II. Moreover, the
dependence on the mass implemented in the Monte Carlo program,
employed to obtain the acceptance, in very mild.
The top pole mass was also extracted from the
measurement of the $t\bar t+1$~jet cross
section, more sensitive to
$m_t$ than the inclusive $t\bar t$ rate \cite{atlttj,cmsttj}.
In Ref.~\cite{ttj}, the NLO $t\bar tj$ cross section was
calculated through the POWHEG-BOX, using the
pole mass, and matched to PYTHIA.
Reference~\cite{fuster} computed instead
the NLO $t\bar tj$ rate 
in terms of the $\overline{\rm MS}$ mass and compared
the result with the LHC measurements:
the values of pole and $\overline{\rm MS}$ masses, extracted by following the
methods in \cite{ttj} and \cite{fuster}, are nonetheless in agreement.

Other proposed methods to reconstruct $m_t$ 
rely on kinematic properties of
top-decay final states.
It was found that the peak of the energy of the $b$-jet
in top decay at LO is independent of the boost 
from the top to the laboratory frame, as well as of the
production mechanism \cite{roberto}.
The CMS Collaboration measured the top mass from the $b$-jet
energy peak data at 8 TeV in \cite{bj}.
The $b$-jet+lepton invariant-mass
($m_{b\ell}$) spectrum was used by CMS to reconstruct 
$m_t$ in the dilepton channel, by comparing the data with
PYTHIA \cite{mbl}.
The endpoints of distributions like 
$m_{b\ell}$, $\mu_{bb}$ and $\mu_{\ell\ell}$, 
where $\mu_{bb}$ and $\mu_{\ell\ell}$
are a generalization of the $b\bar b$ and $\ell^+\ell^-$ invariant masses
in the dilepton channel, $b$ being a $b$-jet in top decay,
were also explored to constrain $m_t$ \cite{end}.
Since $b$-flavoured jets can be calibrated directly
from data, Monte Carlo uncertainties in the endpoints
are mostly due to colour reconnection. 

Finally, purely leptonic observables in the dilepton channel,
such as the Mellin moments of lepton energies or transverse
momenta, 
were proposed to measure $m_t$ as they do not require the
reconstruction of the top quarks \cite{frix}.
Such quantities exhibit pretty small
hadronization effects, but they are sensitive to the
production mechanism, to the Lorentz boost from the top rest frame
to the laboratory frame, as well as 
to higher-order corrections.
Preliminary analyses have been carried out in \cite{cmslep}
(CMS, based on LO MadGraph)
and \cite{nisius} (ATLAS, based on the MCFM NLO parton-level code
\cite{mcfm}) and are expected to be improved by
matching NLO amplitudes with shower/hadronization
generators.

\section{Theory and Monte Carlo uncertainties in the top-mass extraction}

In Ref.~\cite{wave}, where the extraction of the
world average is described, the theory uncertainty accounts for
about 540 MeV of the overall 710 GeV systematics.
In particular, Ref.~\cite{wave} distinguishes the contributions
due to Monte Carlo generators, radiation effects, colour
reconnection and parton distribution functions (PDFs).

The Monte Carlo systematics is due to the differences in
the implementation of parton showers, matrix-element matching,
hadronization and underlying event in the various programs available
to describe top-quark production and decay.
There is no unique way to
estimate this uncertainty, though: one can either compare two
different generators or choose a code and explore how its
predictions fare with respect to variations of the parameters.
For example, in \cite{wave}, CDF compares HERWIG and PYTHIA,
while D0 uses ALPGEN+PYTHIA and ALPGEN+HERWIG \cite{alpgen};
both Tevatron experiments use MC@NLO to gauge the overall
impact of NLO corrections. At the LHC, ATLAS compares
MC@NLO with POWHEG for the NLO contributions and PYTHIA
with HERWIG for shower and hadronization; CMS instead
confronts MadGraph with POWHEG.

The radiation uncertainty gauges the effect of
initial- and final-state radiation on the top mass and is typically
obtained by varying in suitable ranges the relevant parameters 
in the parton-shower generators.
Concerning PDFs, there are  strategies to gauge the
induced error on $m_t$ in the different experiments, although
using two different sets or a given set but with different 
parametrizations are common trends.

Colour reconnection is another source of error on $m_t$,
accounting for about 310 MeV in \cite{wave}:
the very fact that, for example, a bottom 
quark in top decay ($t\to bW$) can be colour-connected to 
an initial-state antiquark does not have its counterpart in
$e^+e^-$ annihilation and therefore its modelling in  
Monte Carlo event generators may need retuning at hadron colliders.
Moreover, this phenomenon is an irreducible uncertainty
in the interpretation of the measured mass as a pole mass.
Investigations on the impact of colour reconnection
on $m_t$
were undertaken in \cite{spyros,corc1},
in the frameworks of PYTHIA and HERWIG, respectively.
In particular, Ref.~\cite{corc1} addresses this
issue by simulating fictitious top-flavoured hadrons in HERWIG
and comparing final-state distributions, such as the $BW$ invariant
mass, with standard $t\bar t$ events. In fact, in the top-hadron case,
assuming $T$ decays according to the spectator model, the $b$ quark
is forced to connect with the spectator or with antiquarks in its 
own shower, namely $b\to bg$, followed by $g\to q\bar q$,
and colour reconnection is suppressed.
Furthermore, the analysis \cite{corc1} may also serve 
to address the relation between the measured
mass, often called `Monte Carlo' mass, with the pole mass, since
the mass of a $T$-hadron can be related to any top-quark mass
definition by means of lattice, potential models or Non Relativistic QCD.

More recently, work has been carried out to assess the
dependence of the top-quark mass, extracted by means
of the Mellin moments ${\cal M}_n$ of some variables
related to $B$-hadrons in top decays,
on the Monte Carlo shower and hadronization parameters \cite{cfk},
extending the investigation in \cite{mescia},  which studied
only the $m_{B\ell}$ quantity.
In fact, when addressing $B$-hadron rather than
$b$-jet observables, one should deal
with fragmentation uncertainties, rather than with the jet-energy scale,
entering in measurements relying on $b$-jets.
If ${\cal M}_1=\langle O\rangle$ is the average value of some
observable $O$ and $\theta$ a generic generator parameter,
one can write the following relations:
\begin{equation}
  \frac{dm_t}{m_t}=\Delta_O^m\  \frac{d\langle O\rangle}{\langle O\rangle}\ \ ;\ \ \frac{d\langle O\rangle}{\langle O\rangle}=
  \Delta_\theta^O\ \frac{d\theta}{\theta}\  \Rightarrow\ 
  \frac{dm_t}{m_t}=\Delta_\theta^m \frac{d\theta}{\theta},
  \end{equation}
where we defined $\Delta_\theta^m=\Delta_O^m\ \Delta_\theta^O$.
Therefore, if one requires, e.g.,  a relative error below $0.3\%$ on
$m_t$, namely $dm_t/dm_t<0.003$, one should also have
$\Delta_\theta^m (d\theta/\theta)<0.003$.
\begin{table*}
  \tiny
\begin{centering}
\begin{tabular}[t]{|c|c|c|c|c|c|c|c|c|c|c|}
\hline 
\multirow{2}{*}{$\mathcal{O}$} & \multirow{2}{*}{$\DeltaMO$} & \multicolumn{9}{c|}{$\Delta_{\theta}^m$}\tabularnewline
\cline{3-11} 
&  & PSPLT(2) & QCDLAM & CLPOW & CLSMR(2) & CLMAX & RMASS(5) & RMASS(13) & VGCUT & VQCUT\tabularnewline
\hline 
\hline 
$m_{B\ell}$ & 0.52 & 0.036(4) & -0.008(2) & -0.007(5) & 0.002(3) & -0.007(4) & 0.058(1) & 0.06(5) & 0.003(1) & -0.003(3)\tabularnewline
\hline 
$p_{T,B}$ & 0.47 & 0.072(1) & -0.03(9) & -0.02(7) & 0.0035(5) & -0.03(5) & 0.11(9) & 0.12(5) & 0.0066(2) & -0.006(5)\tabularnewline
\hline 
$E_{B}$ & 0.43 & 0.069(7) & -0.026(7) & -0.017(5) & 0.0038(9) & -0.01(2) & 0.12(1) & 0.12(2) & 0.006(2) & -0.007(5)\tabularnewline
\hline 
$E_{\ell}$ & 0.13 & 0.0005(5) & -0.04(3) & 0.04(2) & -0.0002(2) & -0.004(4) & 0.008(3) & 0.008(2) & -0.002(5) & 0.008(2)\tabularnewline
\hline 
\end{tabular}
\par\end{centering}
\caption{\label{tabhw}
  Dependence of $m_t$ 
  HERWIG 6 shower and hadronization parameters.}
\end{table*}
In Table~\ref{tabhw}, we present the $\Delta$ factors $\DeltaMO$
and $\Delta_\theta^m$, assuming that the top mass is extracted
in the dilepton channel by
means of the first Mellin moment of observables $O$, like
the $B\ell$ invariant mass, the energy $E_B$ and the transverse momentum
$p_{T,B}$ of $B$ hadrons in top decays, the energy $E_\ell$ of charged
leptons in $W$ decays. The $\theta$ entries in Table~\ref{tabhw}
are parameters of the HERWIG 6 event generator, implementing
the cluster hadronization model.
In detail, PSPLT(2) is a parameter 
ruling the mass distribution
of the decays of $b$-flavoured clusters, while CLMAX(2) and CLPOW
determine the highest allowed cluster mass.
Furthermore, unlike Ref.~\cite{mescia}, 
which just accounted for cluster-hadronization parameters, Ref.~\cite{cfk}
also investigates the dependence of top-quark mass observables on the following
parameters: RMASS(5) and RMASS(13), the bottom and gluon effective
masses, respectively, and the virtuality cutoffs, VQCUT for quarks and VGCUT for
gluons, which are added to the parton masses in the shower.
The impact
of changing QCDLAM, the HERWIG parameter playing
the role of an effective $\Lambda_{\rm QCD}$ in the
shower definition of the strong coupling constant 
\cite{cmw}, is also examined.
Overall, from Table~\ref{tabhw} one learns that, if one aims
at $dm_t/m_t<0.003$, the parameters PSPLT(2), QCDLAM, CLPOW,
CLMAX and the $b$-quark and gluon effective masses are to be known
with a relative precision of 10\%. The dependence of $m_t$ on
CLSMR(2) and the cutoffs VQCUT and VGCUT is instead very mild
and, in principle,
it would be sufficient determining only the order of magnitude of such
parameters to meet a 0.3\% goal on $m_t$.
More details on the dependence of the top-quark mass 
on hadronization and shower parameters will be soon
available in \cite{cfk}.

Another recent investigation on the sensitivity of $m_t$ to the 
Monte Carlo modelling
was carried out in Ref.~\cite{schwartz},
where the authors investigated the PYTHIA uncertainty in $m_t$ in
the lepton+jets channel. It was then found that the error on the
top mass can be significantly reduced if one calibrates the
$W$ mass or applies the soft-drop jet grooming.

On the top of the uncertainties in the $m_t$
determination at the LHC, there are long-standing theoretical issues
affecting the accuracy on the top mass, namely
the interpretation of the measured quantity in terms of the
pole mass and the renormalon ambiguity on the pole mass:
a review on such topics can be found in \cite{corc2}.
In fact, Ref.~\cite{buten} compared PYTHIA with a 
 NLO+NNLL SCET calculation for $e^+e^-\to t\bar t$ annihilation
 and calibrated the top mass used in the computation,
 the so-called MSR definition, 
to agree with the Monte Carlo 2-jettiness distribution.
Within the error range,
the mass parameter in PYTHIA is consistent with the fitted value of 
$m_{\rm MSR}(1~{\rm GeV})$, while 
$(0.57\pm 0.28)$~GeV is the discrepancy with respect to the 
corresponding pole mass.
Ref.~\cite{groom} proposed instead the measurement of $m_t$
in $pp$ collisions by using boosted top jets with light soft-drop grooming,
in lepton+jets and all-jet final states.
The groomed top-jet mass spectrum was then calculated 
resumming soft- and collinear-enhanced contributions 
in the NLL approximation and compared with the spectra yielded
by PYTHIA, trying to calibrate the PYTHIA mass parameter to
reproduce the resummed distribution. The result of this calibration
is that the pole mass is about 400-700 MeV smaller than the
tuned mass in the Monte Carlo generator.

As for the renormalon ambiguity, two recent analyses,
i.e. Refs.~\cite{nasben} and \cite{hpole}, estimated 
the uncertainty in
the top pole mass due to renormalons, obtaining 
about 110 and 250 MeV, respectively: though differing by about a factor
of 2, such results are both smaller than the current error on
the measured top mass.

\section{Conclusions}

I discussed the main strategies to determine the top mass 
at the LHC, emphasizing the role played by the theory uncertainties.
The so-called standard measurements, such as those relying
on template, matrix-element or ideogram methods, are
based on the reconstruction of the top-decay products, and therefore
they yield results close to the top-quark pole mass:
however, a careful exploration of the theory error is mandatory.
In particular, the ongoing work in \cite{corc1},
studying fictitious top-flavoured hadrons, 
should shade light
on both colour reconnection and relation between measured mass
and pole mass. It will be therefore very interesting to compare
the eventual uncertainties according to the method in \cite{corc1}
with the errors obtained in \cite{buten,groom}, comparing resummations and
event generators, as well as with the renormalon ambiguity
gauged in \cite{nasben,hpole}.

Among the alternative measurements, 
extracting $m_t$ by 
confronting the
$t\bar t$ and $t\bar t j$ cross sections with
NLO or NNLO calculations
allows a
clean extraction of the pole mass.
The errors are substantially larger than in the standard
measurements, but nonetheless they are supposed to become
much smaller once the LHC statistics increase.
Other methods, employing endpoints, leptonic observables
or observables like the $b$-jet+lepton mass distribution,
are very interesting and worthwhile to be further developed,
since they are sensitive to different effects with respect to the
standard determination. For example, the endpoint method minimizes the
impact of the Monte Carlo generators, while leptonic quantities do not
need the reconstruction of the top quarks.

Particular care was taken in this talk about the Monte Carlo uncertainty
in the $m_t$ determination, namely the dependence of $m_t$ 
on hadronization parameters, once it is measured from
$B$-hadron observables, such as the $B\ell$ invariant mass
or the $B$-energy or transverse-momentum spectra.
I presented some results yielded by the HERWIG event generator,
showing that most parameters are to be tuned with an accuracy of
at least 10\%, for the sake of meeting a 0.3\% precision goal on
$m_t$. More details on this investigation and on
the dependence of $m_t$ on the parameters of PYTHIA, the other
multi-purpose parton shower generator employed in
the analysis, will be soon available in
\cite{cfk}. It will be challenging comparing the
hadronization uncertainties
obtained in \cite{cfk} with those
relying on NLO+shower generators, such as the recent
$t\bar t$ POWHEG
implementation presented in \cite{powtop},
accounting for non-resonant contributions and
for the interference between
top production and decay. Furthermore, at parton-level,
the full process $pp\to W^+W^-b\bar b\to (\ell^+\nu_\ell)
(\ell^-\bar\nu_\ell)$ was lately computed at NLO and compared
with scenarios where NLO $t\bar t$ production is matched with
different top-decay modelling, namely LO and NLO top decays in 
the narrow-width approximation, as well as parton showers
\cite{gudrun}.
It will be certainly very interesting confronting the approaches in
Refs.~\cite{powtop} and \cite{gudrun}, and the induced error in
the $m_t$ determination.

In summary, the current world-average $m_t$ analysis exhibits
an uncertainty of about 0.5\%
and higher accuracies
are foreseen in the near future, thanks to the large statistics.
Given the relevance of $m_t$ in the
Standard Model, any progress in improving the
present higher-order calculations and Monte Carlo event
generators, for the sake of assessing reliably the  
theoretical error, including the
interpretation of the measurements
in terms of the pole mass, will therefore be especially desirable.


\begin{thebibliography}{99}
\bibitem{gfitter}
  Gfitter Group, Eur. Phys. J. C74 (2014) 3046.
\bibitem{degrassi}
G. Degrassi, S. Di Vita, J. Elias-Mir\'o, J.R. Espinosa, G.F. Giudice, 
G. Isidori and A. Strumia, JHEP 1208 (2012) 098.
\bibitem{wave}
ATLAS and CDF and CMS and D0 Collaborations,  arXiv:1403.4427 [hep-ex].
\bibitem{atlas1}
ATLAS Collaboration, Phys. Lett. B761 (2016) 350.
\bibitem{cms1}
CMS Collaboration, Phys. Rev. D93 (2016) 072004.
\bibitem{herwig}
J. Bellm et al., Eur. Phys. J. C76 (2016) 196.
\bibitem{pythia}
T. Sj\"ostrand et al.,
  Comput. Phys. Commun. 191 (2015) 159.
\bibitem{mcnlo}
J. Alwall et al., JHEP 1407 (2014) 079.
\bibitem{powheg}
S. Alioli, P. Nason, C. Oleari and E. Re, JHEP 1006 (2010) 043.
\bibitem{sigmaatl}
  ATLAS Collaboration, Eur. Phys. J. C74 (2015) 3109.
\bibitem{sigmacms}
CMS Collaboration, JHEP 08 (2016) 029.
\bibitem{alex}
M. Czakon, P. Fiedler and A. Mitov, Phys. Rev. Lett. 110 (2013) 252004.
\bibitem{atlttj}
ATLAS Collaboration, JHEP 1510 (2015) 121.
\bibitem{cmsttj}
CMS Collaboration, CMS-PAS-TOP-13-006.
\bibitem{ttj}
S. Alioli, P. Fernandez, J. Fuster, A. Irles, 
S.-O. Moch, P. Uwer and M. Vos, Eur. Phys. J. C73 (2013) 2438.
\bibitem{fuster}
  J. Fuster, D. Melini, P. Uwer and M. Vos, arXiv:1704.00540 [hep-ph].
\bibitem{roberto}
  K. Agashe, R. Franceschini, D. Kim and M. Schulze, Eur. Phys. J.
  C76 (2016) 636.
 \bibitem{bj}
CMS Collaboration, PoS ICHEP2016 (2016) 743.
  \bibitem{mbl}
CMS Collaboration, CMS-PAS-TOP-14-014.
\bibitem{end}
  CMS Collaboration, Eur. Phys. J. C73 (2013) 2494.
\bibitem{frix}
  S. Frixione and A. Mitov, JHEP 1409 (2014) 012.
\bibitem{cmslep}
  CMS Collaboration, CMS-PAS-TOP-16-002.
\bibitem{nisius}
  ATLAS Collaboration, ATLAS-CONF-2017-044.
\bibitem{mcfm}
J.M. Campbell and R.K. Ellis, J. Phys. G42 (2015) 015005.  
\bibitem{alpgen}
  M.L. Mangano, M. Moretti, F. Piccinini, R. Pittau, A. Polosa,
  JHEP 0307 (2003) 001.
\bibitem{spyros}	
S. Argyropoulos and T. Sj\"ostrand, JHEP 1411 (2014) 043.
\bibitem{corc1}
G. Corcella, EPJ Web Conf. 80 (2014) 00019.
\bibitem{cfk}
  G. Corcella, R. Franceschini and D. Kim, in preparation.
\bibitem{mescia}
  G. Corcella and F. Mescia, Eur. Phys. J. C65 (2010) 171;
Erratum-ibid. C68 (2010) 687.
\bibitem{cmw}
  S. Catani, B.R. Webber, and G. Marchesini, Nucl. Phys. B349
(1991) 635.
\bibitem{schwartz}
A. Andreassen and M.D. Schwartz, JHEP 1710 (2017) 151.
\bibitem{corc2}
G. Corcella,   arXiv:1709.09878 [hep-ph].
\bibitem{buten}
  M. Butensch\"on, B. Denhadi, A.H. Hoang, V. Mateu, M. Preisser and
  I.W. Stewart, Phys. Rev. Lett. 117 (2016) 232001.
\bibitem{groom}
  A.H. Hoang, S. Mantry, A. Pathak and I.W. Stewart, 
 arXiv:1708.02586 [hep-ph].
\bibitem{nasben}
  M. Beneke, P. Marquard, P. Nason and M. Steinhauser, Phys. Lett. B775 (2017)
  63.
\bibitem{hpole}
  A.H. Hoang, C. Lepenik and M. Preisser, JHEP 1709 (2017) 099.
\bibitem{powtop}
  T. Jezo, J.M. Lindert, P. Nason, C. Oleari and S. Pozzorini,
  Eur. Phys. J. C76 (2016) 691.
\bibitem{gudrun}
  G. Heinrich, A. Maier, R. Nisius, J. Schlenk, M. Schulze,
  L. Scyboz and J. Winter, arXiv:1709.08615 [hep-ph].
\end{thebibliography}
\end{document}